\documentstyle[12pt]{article}
\input dspace12
\title{\large \bf Pseudo Dirac Neutrinos in Seesaw model}
\author{Gautam Dutta and Anjan S. Joshipura\\
Theory Group, Physical Research Laboratory\\
Navrangpura, Ahmedabad 380 009, India}
\date{}
\maketitle
\begin{abstract}
Specific class of textures for the Dirac and Majorana mass matrices
in the seesaw model leading to a pair of almost degenerate neutrinos
is discussed. These textures can be obtained by imposing a horizontal
$U(1)$ symmetry. A specific model is discussed in which: (1) All
three neutrino masses are similar in magnitude and could lie around
eV providing hot component of the dark matter in the universe. (2)
Two of these are highly degenerate and their ${\hbox{(mass)}}^2$
difference could solve the solar neutrino problem through large angle
MSW solution. (3) The electron neutrino mass may be observable
through Kurie plot as well as through search of the neutrinoless
double beta decay.
\end{abstract}
PACS $\#$ $14.60.Pq$, $13.10.+q$, $13.55.+d$, $14.60.Gh$
\newpage

\begin{document}
\middlespace

\noindent
\section{Introduction}
\noindent Variety of experimental findings strongly suggest that
possibly \cite{asj} all the neutrinos are massive. But these masses
have to be much smaller than the masses of other fermions. This
smallness of neutrino mass is theoretically well-understood in the
framework of seesaw mechanism \cite{GRSY}. Within this framework,
neutrino masses are described by the following mass matrix in the
basis $(\nu^c_{iL},\nu_{iR})$
\begin{equation}
{\cal M}_{\nu} = \left[\begin{array}{ccc} 0 & m_D\\
                         m_D^T & M_R \end{array} \right] \label{mnu}
\end{equation}
$m_D$ and $M_R$ are $3\times 3$ matrices in the generation space
labeled by $i=1,2,3$. When the scale associated with $M_R$ is much
higher than that of $m_D$, the masses of the light neutrinos are
described by the effective matrix
\begin{equation}
m_{eff.}=-m_D M_R^{-1} m_D^T                       \label{meff}
\end{equation}
The overall magnitude of $M_R$ is determined either by the grand
unification scale or by some intermediate symmetry breaking scale in
grand unified theories \cite{seesaw,langacker}. The Dirac mass matrix
$m_D$ arise in the same way as other fermion masses. Its eigenvalues
are therefore expected to be similar to other fermion masses. These
two aspects together lead to very light neutrinos through $m_{eff.}$.
The exact pattern of neutrino masses depends upon the structures of
$m_D$ and $M_R$. The $m_D$ can be related to other fermion  mass
matrix, typically up-quark, in $SO(10)$ models but $M_R$ remains
unconstrained. In the absence of any knowledge on $M_R$, it is
usually chosen \cite{seesaw,langacker} to be proportional to
identity. In this case, eq.(\ref{meff})
immediately implies that the light neutrino masses display
strong hierarchy similar to other fermion masses. Typically
\cite{seesaw,langacker} for $M_R\sim$GUT scale $\sim 10^{16}GeV$
one has $m_{\nu_e} \sim 10^{-11}eV$, $m_{\nu_\mu} \sim 1.5\times
10^{-7}eV$ and $m_{\nu_\tau} \sim 10^{-3}eV$ while for $M_R$ in the
intermediate range $\sim 10^{12}GeV$, $m_{\nu_e} \sim 10^{-7}eV$,
$m_{\nu_\mu} \sim 1.5\times 10^{-3}eV$ and $m_{\nu_\tau} \sim 10eV$.
Immediate consequence of these values is that at most $\nu_\tau$ can
have mass in the eV range if $M_R \ge 10^{12}GeV$. Thus one cannot
hope to see electron neutrino mass through direct laboratory
experiments. It is important to realize that the above masses, many
times taken \cite{seesaw,langacker} as ``predictions'' of seesaw
model, crucially depend upon $M_R$ being proportional to identity.
One may naively expect that even if $M_R$ has some structure, as long
as its non-zero elements (more precisely eigenvalues \cite{valle}) do
not display any hierarchy, the predictions would remain true, at
least qualitatively. These naive expectations need not always hold if
some symmetry enforces a texture for $m_D$ and/or $M_R$. The purpose
of this paper is to present examples of different structures for
$M_R$ which display completely different hierarchy in neutrino masses
without making any unnatural assumptions on values of parameters
entering $m_D$ and $M_R$. In the examples to be presented, two of the
neutrinos turn out to have almost equal but opposite eigenvalues.
These therefore form a pseudo Dirac neutrino \cite{wolfenstein}.
These examples can be realized through explicit models and
we discuss a specific model in which all three neutrinos have similar
masses whose values could be $O(eV)$. Two of them are highly
degenerate and have typical mass square difference of
$O(10^{-5}eV^2)$ needed for solution to the solar neutrino problem
through Mikheyev Smirnov Wolfenstein (MSW) mechanism.

In the next section, we present an explicit model with pseudo Dirac
neutrinos. The expected patterns of neutrino masses and mixing as
well as their implications are discussed in section 3 and conclusions
are presented in section 4. Appendix gives a class of textures for
$M_R$ which are similar to the one discussed in text and which lead
to pseudo Dirac neutrinos.

\noindent
\section{Pseudo Dirac neutrinos}
We present here a specific structure for $m_D$ and $M_R$ which
lead to pseudo Dirac neutrinos. These structures can be easily
obtained by imposition of a horizontal $U(1)$ symmetry. Horizontal
$U(1)$ symmetries have been used long ago in order to constrain the
quark \cite{wali} as well as neutrino \cite{asj1} masses. These have
also been studied recently \cite{babu} with a view of obtaining large
mixing among neutrinos required in order to solve the atmospheric
neutrino problem \cite{asj,seesaw}. The $U(1)$ assignments that we
employ here and the resulting structure for the neutrino masses is
somewhat different from the one considered in refs.\cite{asj1,babu}.
We shall require $U(1)$ to be vectorial and assume that the $i^{th}$
generation of leptons carry the $U(1)$ charge $X_i$ and that no two
$X_i$ are identical. The ordinary Higgs doublet $\Phi$ is assumed to
be neutral under the $U(1)$ symmetry. As an immediate consequence,
both the charged leptons as well as Dirac mass matrix $m_D$ in
eq.(\ref{mnu}) are forced to be diagonal. The Majorana mass term can
still have texture. We are interested in the Fritzsch type
\cite{fritzsch} of structures for $M_R$. This can be easily obtained
by introducing an $SU(2) \otimes U(1)$ singlet field $\eta$ with
non-trivial $U(1)$ charge $X$. The $M_R$ receives contributions from
the following terms:
\begin{equation}
-{\cal L}_R = {1 \over 2}\nu^T_{Ri}(M_{ij}+ \Gamma_{ij} \eta +
     \Gamma'_{ij}\eta^* )C \nu_{Rj} + h.c     \label{lag}
\end{equation}
All the three terms are possible if total lepton number is not
assumed to be conserved \cite{asj2}.

Now consider the specific assignment
\begin{equation}
\begin{array}{rclcl}
X_2 & = & -{1\over 3}X_3 & = & -{1\over 2}X \\
X_1 & = & -{5\over 3}X_3 &   &
\end{array}    \label{sp}
\end{equation}
with $X_3 \ne 0$. Then $M_{ij}$ in eq.(\ref{lag}) are zero for all
$i$ and $j$. Moreover only $\Gamma_{13} = \Gamma_{31}$, $\Gamma_{22}$
and $\Gamma'_{23} = \Gamma'_{32}$ are allowed to be non-zero. This
leads to the following texture for $M_R$:\\
\begin{equation}
M_R = \left[\begin{array}{ccc} 0 & 0 & M_1 \\
                       0 & M_3 & M_2 \\
                       M_1 & M_2 & 0 \end{array} \right]   \label{mr}
\end{equation}   \\
$M_1 =\Gamma_{13} \langle \eta \rangle, M_3=\Gamma_{22} \langle \eta
\rangle$ and $M_2 =\Gamma'_{23} \langle \eta^* \rangle$.
If we denote the elements of the diagonal matrix $m_D$ by
$m_i$($i=1,2,3$) then the effective mass matrix for the light
neutrinos is given by:\\
\begin{equation}\\
m_{eff.}=-m_D M_R^{-1} m_D^T =
-\left[\begin{array}{rrr}
m_1^2 M_2^2 & -M_1 M_2 m_1 m_2 & M_1 M_3m_1 m_3\\
-M_1 M_2 m_1 m_2 & M_1^2 m_2^2 &0   \\
M_1 M_3 m_1 m_3 & 0 & 0 \end{array} \right] {1 \over M_1^2 M_3}
						\label{meff1}
\end{equation}\\
We shall assume parameters $M_{1,2,3}$ to be similar in magnitude
(often we will take them identical for some of the numerical
estimates). In addition we will also assume hierarchy in the Dirac
masses $m_1 << m_2<< m_3$. Both the above assumptions are
natural assumptions made in the usual seesaw picture
\cite{seesaw,langacker}. But since $M_R$ is different from identity,
the resulting pattern of neutrino masses is completely different from
the usual seesaw predictions. The eigenvalues of $m_{eff.}$ are given
as follows: \\
\begin{eqnarray}
m_{\nu_1} & \approx &-{m_1m_3\over M_1}\left\{1 + {1\over
               2}{\epsilon\over \delta -1}\right\} \nonumber \\
m_{\nu_2} & \approx &-\frac{m_2^2}{M_3}\left\{1 + \frac{\epsilon}{1 -
\delta^2 }\right\} \nonumber \\
m_{\nu_3} & \approx &\frac{m_1m_3}{M_1}\left\{1 + {1\over
               2}\frac{\epsilon}{\delta +1}\right\}
       \label{ev}
\end{eqnarray} \\
where
\begin{eqnarray}
\epsilon &\equiv& \left({m_1\over m_2}\right)^2\left({M_2\over
M_1}\right)^2 \nonumber\\
\delta &\equiv& {m_1 m_3 \over m_2^2}\left({M_3 \over M_1}\right)
\label{epsidel}
\end{eqnarray}\\
The parameter $\epsilon << 1$ with the above stated assumptions
while $\delta$ could be $O(1)$. In the $\epsilon \rightarrow 0$
limit, two of the neutrinos are exactly degenerate while the presence
of the $\epsilon$ term introduces small splitting with the
${\hbox{(mass)}}^2$ difference \\
\begin{equation}
\Delta_{31} \approx 2\left({m_1 m_3 \over M_1} \right)^2 {\epsilon
\delta \over \delta^2 -1}                    \label{del31}
\end{equation}\\
Hence for $\epsilon << 1$, these neutrinos form a pair of pseudo
Dirac particles \cite{wolfenstein}. The occurrence of a pseudo Dirac
neutrino here is somewhat of a surprise. One would have expected it
\cite{branco} if both $m_D$ and $M_R$ possessed some approximate
global symmetry. For example, if $M_2$ is taken to be small, $M_R$ is
approximately invariant under $L_e-L_\tau$ and one could have
expected a pseudo Dirac neutrino. But $M_R$ in eq.(\ref{mr}) does not
possess any approximate global symmetry as long as $M_1\sim M_2 \sim
M_3$. Despite this, the hierarchy in $m_i$ (combined with
specific texture for $M_R$) makes $m_{eff.}$ approximately invariant
under $L_e-L_\tau$ symmetry resulting in a pseudo Dirac neutrino. In
practice $\epsilon$ could be quite small, e.g. if $m_1(m_2)$ is
identified with $m_u(m_c)$ then $\epsilon \sim 10^{-5}$ for $M_2 \sim
M_1$. Thus degeneracy of two neutrinos is expected to be quite good.

The mixing among neutrinos implied by eq.(\ref{meff1}) can be easily
worked out for $\epsilon << 1$. In general, the eigen vectors of
$m_{eff.}$ are given by
\[
\Psi_i = N_i \left( \begin{array}{c} x_i\\
			  y_i\\
			  z_i
	        \end{array} \right)
\]
with
\begin{eqnarray}
x_i&=&\frac{m^0_{\nu_2} - m_{\nu_i}}{B} y_i \nonumber\\
z_i&=&-{m^0_{\nu_1} \over m_{\nu_i}} x_i \label{wfc}
\end{eqnarray}\\
where
\[
m^0_{\nu_1}\equiv -{m_1 m_3 \over M_1} \equiv  -m^0_{\nu_3}; \;\;
m^0_{\nu_2} \equiv
{m^2_2 \over M_3}\;\; {\hbox{and}}\;\; B\equiv {m_1 m_2 M_2 \over M_1
M_3}.
\]
With $m_{\nu_i}$ given in eq.(\ref{ev}), one could determine $\Psi_i$
and
hence the mixing angles. The three wavefunctions are approximately
given by \\
\[
\Psi_1 \approx {1 \over \sqrt{2}}\left(\begin{array}{c}
1\\0\\1\end{array}\right)\;\;
\Psi_2 \approx \left(\begin{array}{c} 0\\1\\0\end{array}\right)\;\;
{\hbox{and}}\;\;
\Psi_3 \approx {1 \over \sqrt{2}}\left(\begin{array}{c}
1\\0\\-1\end{array}\right)\;\;
\]
\\
$\Psi_1$ and $\Psi_3$ are maximally mixed to form a pseudo Dirac
neutrino. Deviation of this mixing angle from $45^0$ is very
small. Using eqs.(\ref{wfc}) and (\ref{ev}), we see that \\
\begin{equation}
\tan\theta_{13}\approx 1 - {1\over 2}{\epsilon \over \delta -1}
						\label{tan13}
\end{equation} \\
The angle $\theta_{13}$ represents mixing among $\nu_e$ and
$\nu_\tau$ states produced in association with $e$ and $\tau$
respectively if the charged lepton mass matrix is diagonal as is the
case here. This has important implications for the solar neutrino
problem as we will see in the next section.

$M_R$ of eq.(\ref{mr}) possesses a generalized Fritzsch type
\cite{fritzsch}
structure with one pair of off-diagonal and two diagonal elements
being zero. One could write down other similar structures. Just as in
the example studied here, most other similar structures admit
almost degenerate neutrinos in the limit $m_1<<m_2<<m_3$. We list
these structures separately in the Appendix.

\noindent
\section{Phenomenological Implications}
We shall now investigate the implications of the specific structures
for the neutrino masses and mixing derived in the last section. These
clearly depend upon the unknown values of the Dirac masses $m_i$. To
be specific, we shall assume these masses to coincide with the
up-quark masses. Moreover, we shall assume $M_1=M_2=M_3$ and denote
them by a common scale $M$. The parameters $\epsilon$, $\delta$ and
$M$ then determine neutrino masses and mixing. It follows from
eq.(\ref{epsidel}) that\\
\begin{eqnarray}
\delta &\equiv& \left({m_u m_t\over m^2_c}\right)\left({M_3\over
M_1}\right)\approx
{m_{\nu_e}\over m_{\nu_\mu}}\approx 0.4{\hbox{-}}0.8 \nonumber\\
\epsilon &\approx & \left({m_u \over m_c}\right)^2 \approx 4\times
10^{-5}\nonumber
\end{eqnarray}
\\
where we have chosen $m_u=10MeV$ , $m_t=100$-$200GeV$ and
$m_c=1.5GeV$. Also from eq.(\ref{ev}), we have $|m_{\nu_e}|\sim
|m_{\nu_\tau}|$.
Hence independent of the numerical value of $M$, all three neutrino
masses are expected to be of the same order. This has to be
contrasted with the conventional seesaw prediction
\[
m_{\nu_e}:m_{\nu_\mu}:m_{\nu_\tau}=m^2_u:m^2_c:m^2_t
\]
obtained with similar assumptions on parameters but with $M_R$
proportional to identity. The common mass of all three neutrinos
would lie in the range ($10^{-7}$-$1eV$) for the Majorana mass scale
$M\sim(10^{16}$-$10^9)GeV$. Hence for values of $M$ in the
intermediate range $\sim10^9GeV$, all three neutrinos would have
masses in the eV range. These neutrinos can together provide the
necessary hot component of the dark matter \cite{hot} which requires
$\sum m_\nu = 7eV$. Moreover, such neutrino spectrum could have
observable consequences for laboratory experiments as well. Note that
two of the neutrinos are highly degenerate. Their ${\hbox{(mass)}}^2$
difference is given by eq.(\ref{del31})\\
\begin{equation}
\Delta_{31}\approx 2 (m_{\nu_e})^2 {\epsilon \delta \over \delta^2 -
1}                                                \label{del312}
\end{equation}\\
It follows that if $m_{\nu_e} \sim m_{\nu_\tau}\sim O(eV)$ then their
${\hbox{(mass)}}^2$ difference is naturally expected to be around
$\sim 10^{-5} eV^2$. This value falls in the range required to solve
the solar neutrino problem through Mikheyev-Smirnov-Wolfenstein
\cite{MSW} mechanism. Thus one can solve the solar
neutrino problem and at the same time obtain an electron neutrino
with mass in the observable range unlike in the seesaw models
considered so far in the literature \cite{seesaw,langacker}.

The detailed analysis of the four solar neutrino experiments
constrain the mixing angle as well \cite{smirnov}. The mixing angle
between
$\nu_e$-$\nu_\tau$ is predicted to be large in our case. It turns out
in fact to be too large to be consistent with observations if charged
leptons do not mix among themselves. If the vacuum mixing angle is
close to $\pi/4$ then the survival probability for $\nu_e$ is
independent of energy. Such an energy independent survival
probability
is not favored when the results of all four experiments are
combined. They do allow large angle solution but $\sin^2 2\theta$ (in
our case $\theta\equiv\theta_{13}$) is required to be \cite{smirnov}
\[
\sin^2 2\theta_{13} < 0.85
\]
This constrain is not satisfied by the angle $\theta_{13}$ of
eq.(\ref{tan13}).
$\theta_{13}$ represents the mixing between physical
$\nu_e$-$\nu_\tau$ states if the charged lepton mass matrix is
diagonal as is the case in section 2. The correction to
$\theta_{e\tau}\equiv \theta_{13}$ is proportional to
$\epsilon \sim 10^{-5}$ and is too small to cause significant
deviation from $\pi/4$.
Hence, one must have contribution from the charged lepton mixing in
order to obtain MSW solution consistently. This needs enlargement of
the model. For example, consider adding two more Higgs doublets
$\Phi'_{1,2}$ with
$U(1)_X$ charges 0 and $-{8\over 3}X_3$ respectively to the model
presented in the last section. With a suitable discrete
symmetry ($\Phi'_{1,2}\rightarrow
-\Phi'_{1,2};\; e_R\rightarrow -e_R$), $\Phi'_{1,2}$ can be made to
contribute
only to the charged lepton masses. These are now described by a
mass matrix \\
\begin{equation}
M_l=\left[\begin{array}{ccc} m_{ee}&0&m_{e\tau}\\
		  0&m_\mu &0\\
		  0&0&m_{\tau \tau} \end{array}\right]   \label{ml}
\end{equation}
\\
The neutrino mass matrix $m_D$ and hence $m_{eff.}$ remains the same
as before.
Because of the structure for $M_l$ in eq.(\ref{ml}), the
effective mixing angle
describing $\nu_e$-$\nu_\tau$ mixing is now given by
\[
\theta_{e\tau} \approx \theta_{13}-\phi
\]
with
\[
\tan 2\phi={2m_{e\tau} \over m_{ee}^2+m_{e\tau}^2-m_{\tau\tau}^2}
\]\\
Due to contribution from $\phi$, $\theta_{e\tau}$ need not be very
close to $\pi /4$. The large angle MSW solution typically needs
\cite{MSW}
$\sin^2 2\theta_{e\tau} \approx 0.65$-$0.85$. With $\theta_{13} \sim
45^0$,
this translates to $\phi \sim 10^0$-$20^0$.

The present model makes a definite prediction for the neutrinoless
$\beta\beta$ decay. The amplitude for this process is related to the
({\em 11}) element of the neutrino mass matrix in the basis in which
charged lepton mass matrix is diagonal \cite{seesaw}. It follows
therefore from eq.(\ref{meff1}) that in the model of the earlier
section,
neutrinoless $\beta\beta$ decay amplitude is proportional to
\\
\[
{m_1^2\over M_3}\left({M_2\over M_1}\right)^2 \approx \epsilon
m_{\nu_2}
\]
\\
Hence unless $m_{\nu_2}$ is very large $\sim10^5eV$, the $\nu$-less
$\beta\beta$ decay is not observable. If the charged lepton mass
matrix is non-diagonal as is required here in order to obtain the
right amount of mixing for the MSW solution to work then the
$\nu$-less $\beta\beta$ decay amplitude also changes and is now
proportional to\\
\[
\cos^2\phi \epsilon m_{\nu_2} - 2 \sin\phi \cos\phi m_{\nu_1}
\]\\
with $m_{\nu_2}$ and $m_{\nu_1}$ given by eqs.(\ref{ev}). For $\phi
\sim
10^0$-$20^0$ and $m_{\nu_1}\sim m_{\nu_2}$ this corresponds to
$\sim(0.3$-$0.6)m_{\nu_1}$. Hence, the $\nu_e$ mass $\sim 2eV$ could
lead to
an observable signal in the $\nu$-less $\beta\beta$ decay
\cite{nuless}.
With $m_{\nu_1}\sim m_{\nu_2}\sim m_{\nu_3}\sim 2eV$, one can also
obtain the $\sim 7eV$ needed for obtaining hot component in the dark
matter \cite{hot}.
We have concentrated here on a specific structure among various
possibilities (Appendix) that lead to pseudo Dirac neutrinos. The
quantitative consequences of other structures could be considerably
different. Moreover, within the specific texture, we have assumed
$m_{1,2,3}$ to coincide with $m_{u,c,t}$. Such an identification need
not hold \cite{hsym}. But the qualitative conclusions, namely the
occurrence  of pseudo Dirac neutrino due to approximate $L_e-L_\tau$
symmetry of $m_{eff.}$ is more general and holds as long as
$m_1<<m_2<<m_3$ and $M_1\sim M_2\sim M_3$. This is a naturally
expected pattern even if $m_{1,2,3}$ do not exactly coincide with
$m_{u,c,t}$. Interesting predictions discussed above still remain
true if $m_{1,2,3}$ are chosen to coincide with the corresponding
charged lepton masses instead of the up-quark masses. Now $\epsilon
\sim (m_e/m_\mu)^2$ and $\delta \sim m_e m_\mu /m_\tau^2$. Hence
$\Delta_{31}$ given in eq.(\ref{del312}) now becomes
\[
\Delta_{31}\sim {1\over 8}(m_{\nu_e})^2 \times 10^{-6}
\]
Hence for $m_{\nu_e} \sim 1eV$, one can still solve the solar
neutrino problem through MSW mechanism. $m_{\nu_e} \sim {m_e
m_\tau\over M}$ falls in the eV range if $M\sim 10^{12}GeV$.

\noindent
\section{Conclusions}
Seesaw model as conventionally analyzed generally lead to
hierarchical neutrino masses. In particular, if the Majorana masses
of the right handed neutrinos are large ($>10^9GeV$) then at most
$\nu_\tau$ could have mass around eV range and $\nu_e$ is not
expected to have mass near its laboratory limit. We have presented
and discussed specific textures for neutrino masses which lead to
very different predictions. A particular model analyzed in detail has
all three neutrino masses in the eV range. Hence the model can
naturally provide hot component in the dark matter of the universe.
Two of the neutrinos are nearly degenerate and their
${\hbox{(mass)}}^2$ difference could naturally be in the range
required to solve the solar neutrino problem. In addition, this model
is capable of producing $\nu_e$ mass which could be observed directly
in Kurie plot or as well as through $\nu$-less $\beta\beta$ decay.
The atmospheric neutrino problem \cite{seesaw} cannot be solved in
this model easily since the simultaneous solution of the solar $\nu$,
atmospheric $\nu$ and dark matter problem needs nearly degenerate
neutrinos \cite{asj}. Interesting aspect of the model worth
reemphasizing here is the fact that (near) degeneracy of two of the
neutrinos result here in spite of the fact that $M_R$ does not
possess any global symmetry. The hierarchy in eigenvalues of $m_D$
and texture of $M_R$ conspire to make $m_{eff.}$ approximately
invariant under a global $U(1)$ symmetry resulting in almost
degenerate pseudo Dirac neutrinos. This feature is shown to follow if
$m_D$ is diagonal and $M_R$ has a generalized Fritzsch structure.
Both these can be enforced by a global $U(1)$ symmetry. The study
made here highlights the fact that the seesaw model can  accommodate
a completely different pattern of neutrino masses which is not
thought to be among the conventional predictions of the scheme.
\newpage
\noindent
\begin{center}
{\bf {\LARGE Appendix}}
\end{center}

\noindent
In this appendix we give different structures of the Majorana mass
matrix having generalized Fritzsch structure. These lead to a pair of
almost degenerate light neutrinos when $m_D$ in eq.(1) is diagonal,
i.e. $m_D={\hbox{diag}}(m_1,m_2,m_3)$ as in the text. The following
structures are possible for $M_R$ if det$M_R$ is required to be
non-zero \cite{valle}.
\[
\left[\begin{array}{ccc} 0 & \times & 0 \\
          \times & 0 & \times\\
	    0 & \times & \times \end{array}\right]
          (a1);\hspace{0.5in}
\left[\begin{array}{ccc} 0 & \times & 0 \\
          \times & 0 & \times\\
	    0 & \times & \times \end{array}\right]
          (a2);\hspace{0.5in}
\left[\begin{array}{ccc} 0 & 0 & \times \\
           0 & \times & \times\\
	    \times & \times & 0\end{array}\right]   (a3)
\]
\\
\[
\left[\begin{array}{ccc} 0 & \times & \times \\
          \times & \times & 0\\
	  \times & 0 & 0 \end{array}\right]   (a4);\hspace{0.5in}
\left[\begin{array}{ccc} \times & \times & 0 \\
          \times & 0 & \times\\
	    0 & \times & 0 \end{array}\right]   (a5);\hspace{0.5in}
\left[\begin{array}{ccc} \times & 0 & \times\\
          0 & 0 & \times\\
	    \times & \times & 0 \end{array}\right]   (a6)
\]
\\
Here $\times$ denotes a non-zero entry. Any of these structures can
be obtained by imposing a $U(1)$ symmetry similar to the one
considered in the text. If all the entries in a given $M_R$ are
assumed to be identical (and denoted by $M$) then eigenvalues
$\lambda_i$($i=1,2,3$) of $m_{eff.}$ satisfy the following
characteristic equations: \\
\begin{eqnarray}
\lambda_1 \lambda_2 \lambda_3 & = & {m^2_i m^2_j m^2_k \over M^3}\\
\lambda_1\lambda_2 + \lambda_1\lambda_3 + \lambda_2\lambda_3 & =&
{m^2_j m^2_k \over M^2}\\
\lambda_1 +\lambda_2 +\lambda_3 & = & {(-m^2_p-m^2_i) \over M}
					\label{det}
\end{eqnarray}
\\
with $i\ne j\ne k $ and $p$ either $j$ or $k$. In the absence of
${m_p^2 \over M}$ in eq.(\ref{det}), the above eqs. are solved by the
eigenvalues, ${m_jm_k \over M},{-m_jm_k \over M}$ and ${-m_1^2\over
M}$. Hence as long as $m_p^2$ term represents a small correction, one
would get a Pseudo Dirac neutrino. This naturally depends upon the
exact value of the masses $m_{1,2,3}$. If $m_{1,2,3}$ are identified
with $m_{u,c,t}$ (or $m_{e,\mu,\tau}$), the $m_p^2$ term amounts to a
small corrections and one would obtain Pseudo Dirac neutrino in all
cases except $(a4)$ and $(a5)$.


\begin{thebibliography}{9}
\bibitem{smirnov}A recent review can be found in Smirnov, ref.[1]
\end{thebibliography}
\end{document}